\documentstyle[prl,aps,multicol,eqsecnum,tighten,epsfig,array]{revtex}

\def\ga{\mathrel{\mathpalette\fun >}}
\def\fun#1#2{\lower3.6pt\vbox{\baselineskip0pt\lineskip.9pt
\ialign{$\mathsurround=0pt#1\hfill##\hfil$\crcr#2\crcr\sim\crcr}}}

\def\rn{}
\def\nn#1 #2{#2. #1}                            
\def\nnn#1 #2 #3{#2. #3. #1}                    
\def\nnnn#1 #2 #3 #4{#2. #3. #4. #1}            
\def\nnnnn#1 #2 #3 #4 #5{#2. #3. #4 #5. #1}     

\def\rf#1;#2;#3;#4;#5 {{\frenchspacing\par\rn#1, #3 {\bf #4}, #5 (#2). \par}}
\def\rfbook#1;#2;#3;#4;#5 {{\frenchspacing\par\rn#1, {\it #3} (#5, #4, #2).\par}}
\def\rfprep#1;#2;#3 {{\par\frenchspacing\rn#1, #3 (#2).\par}}

\begin{document}

\twocolumn[
\hsize\textwidth\columnwidth\hsize\csname@twocolumnfalse\endcsname

\title{Complex   dynamics   in  a  simple   model  of   pulsations   for
       Super-Asymptotic Giant Branch Stars}

\author{Andreea Munteanu$^{\rm a}$,
        Enrique Garc\'\i a-Berro$^{\rm a,b}$,
        Jordi Jos\'e$^{\rm c,b}$, \&
        Emilia Petrisor$^{\rm d}$}

\address{$^{\rm  a}$  Departament de F\'{\i}sica  Aplicada,  Universitat
        Polit\`ecnica  de Catalunya,  Jordi Girona  Salgado s/n, M\`odul
        B--5, Campus Nord, 08034 Barcelona, Spain}
\address{$^{\rm  b}$ Institut d'Estudis  Espacials de Catalunya, Edifici
        Nexus, Gran Capit\`a 2-4, 08034 Barcelona, Spain}
\address{$^{\rm  c}$  Departament de  F\'{\i}sica i Enginyeria  Nuclear,
	Universitat   Polit\`ecnica   de   Catalunya,   Av.  V\'\i  ctor
	Balaguer, s/n, 08800, Vilanova i la Geltr\'u (Barcelona), Spain}
\address{$^{\rm   d}$   Departamentul   de   Matematica,   Universitatea
	Politehnica  Timisoara,  Pta  Regina  Maria  1, 1900  Timisoara,
	Romania}

~~

\date{\today}

\maketitle

\medskip

\begin{abstract}

When  intermediate  mass stars reach their last stages of evolution they
show pronounced  oscillations.  This phenomenon happens when these stars
reach the so-called  Asymptotic Giant Branch (AGB), which is a region of
the  Hertzsprung-Russell  diagram  located  at about the same  region of
effective  temperatures but at larger luminosities than those of regular
giant  stars.  The period of these  oscillations  depends on the mass of
the star.  There is growing evidence that these  oscillations are highly
correlated  with mass loss and  that, as the mass  loss  increases,  the
pulsations  become  more  chaotic.  In this  paper  we  study  a  simple
oscillator  which  accounts for the observed  properties of this kind of
stars.  This  oscillator was first proposed and studied in  \cite{IFH92}
and we extend  their study to the region of more  massive  and  luminous
stars --- the region of Super-AGB  stars.  The oscillator  consists of a
periodic  nonlinear  perturbation of a linear  Hamiltonian  system.  The
formalism  of  dynamical  systems  theory has been used to  explore  the
associated  Poincar\'e map for the range of parameters  typical of those
stars.  We have studied and characterized the dynamical behaviour of the
oscillator  as the  parameters  of the model are  varied,  leading us to
explore a sequence  of local and  global  bifurcations.  Among  these, a
tripling  bifurcation  is  remarkable,  which allows us to show that the
Poincar\'e map is a nontwist area  preserving  map.  Meandering  curves,
hierarchical-islands  traps and sticky  orbits also show up.  We discuss
the  implications  of the  stickiness  phenomenon  in the  evolution and
stability of the Super-AGB stars.

\end{abstract}


\draft


]

\section{Introduction}

The  study  of  pulsating   stars  has  attracted  much  attention  from
astronomers.  Pulsational  instabilities  are  found in many  phases  of
stellar  evolution,  and also for a wide range of  stellar  masses  (see
\cite{GS96}   for   an   excellent   review).   Moreover,    pulsational
instabilities provide a unique opportunity to learn about the physics of
stars  and  to  derive  useful   constraints  on  the  stellar  physical
mechanisms that would not be accessible otherwise.  Within the theory of
stellar pulsations, there are three basic characteristics of the motions
that the  associated  model may include or not, namely the  oscillations
can be linear or  nonlinear,  adiabatic  or not, and radial or nonradial
\cite{C80}.  Actual pulsations of real stars must certainly involve some
degree of  nonlinearity  \cite{B93,GS95,BEA01}.  In fact, the  irregular
behaviour  observed in many variable stars is, definitely, the result of
those  nonlinear  effects  \cite{PEA96}.  Moreover,  the  fact  that the
observed  pulsation  amplitudes of variable stars of a given type do not
show huge  variations  from star to star  suggests  the  existence  of a
(nonlinear)  limit-cycle type of behaviour \cite{P90}.  However the full
set of nonlinear equations is so complicated that there are no realistic
stellar  models  for  which  analytic   solutions  exist  and  thus  the
investigations  of  nonlinear  pulsations  rely on  numerical  analysis.
Accordingly,  most  recent  theoretical  studies of  stellar  pulsations
proceed   either   through   pure   numerical    hydrodynamical    codes
\cite{HEA98,H99,WEA00,SEA01,BEA00}  or, conversely,  they adopt a set of
simplifying   assumptions   in  order   to  be  able  to  deal   with  a
extraordinarily complex problem.  Most of these simple models of stellar
pulsation are based on a one-zone type of model which may be  visualized
as a single,  relatively  thin,  spherical  mass shell on top of a rigid
core.  These models have helped  considerably in clarifying  some of the
complicated  physics involved in stellar  pulsations and the role played
by different physical mechanisms.

AGB stars are defined as stars that develop  electron  degenerate  cores
made of matter  which  has  experienced  complete  hydrogen  and  helium
burning,  but not  carbon  burning.  More  luminous  stars  with  highly
evolved  cores are called  Super-AGB  stars and their  core is made of a
mixture of  oxygen-neon  \cite{REA96,GBEA97}.  It is a well known result
that the observational counterparts of Super-AGB stars --- the so-called
Long Period Variables  (LPVs) --- are radial  pulsators.  Moreover, long
term  photometry  of these stars has shown that their light  curves (the
variation of the luminosity with time) usually have some  irregularities
\cite{B98,BEA99}  which lead to a high degree of  impredictability.  Due
to  the  lack  of  appropriate   tools  for  analyzing  these  irregular
fluctuations  of the  luminosity or the stellar  radius, the  scientific
community did not pay much attention,  until  recently, to this category
of  variable  stars.  The  development  of  new  nonlinear   time-series
analysis  tools during the last decade has changed  this  situation.  In
particular, it has been found that these tools have rich applications in
a broad range of astrophysical  situations,  which include time analysis
of gamma-ray  bursts  \cite{NEA94},  of  gravitationally  lensed quasars
\cite{H92}  or  X-rays  within  galaxy  clusters  \cite{SEA94},  and  of
variable white dwarfs  \cite{GEA91}.  These tools have also been used to
analyze theoretical models of stellar pulsations.  In particular, it has
been proven that numerical  hydrodynamical  models  display  cascades of
period doublings  \cite{BK87,KB88,A90}  as well as tangent  bifurcations
\cite{BGK87,A87}.  To  be  more   specific,   it  has   been   confirmed
\cite{SEA96}  that the irregular  pulsations  of W~Vir models are indeed
chaotic and,  furthermore,  it has been shown that the  physical  system
generating  the time series is equivalent to a system of three  ordinary
differential  equations.  A similar  approach has been used also for the
study of the  pulsations  of other types of stars like, for instance two
RV Tau stars,  R~Scuti  \cite{BEA95}  and AC~Her  \cite{KEA98},  and has
provided  significant results concerning the underlying  dynamics.  This
result is not a trivial one since it is not obvious at all that  stellar
pulsations  can be fully  described by such a simple  system of ordinary
differential  equations.  This  stems  from the fact that the  classical
method to physically  describe stellar pulsations is based on the use of
a hydrodynamical code where the partial differential  equations of fluid
dynamics are replaced by a discrete approximation consisting of $N$ mass
shells.  Therefore,   a  set  of   $3N$   coupled   nonlinear   ordinary
differential  equations  must be solved,  where $N$ is typically  of the
order of 60.  Thus,  simple  models do not only help in  clarifying  the
basic behaviour of stellar pulsations but, actually, they may be able to
reproduce with a reasonable accuracy the oscillations of real stars.

Several   such   simple   models   have  been   proposed   and   studied
\cite{B66,BMS66,RR70,S72,BR82,TT88,STT89,UX93}   including  the  one  in
\cite{IFH92}.  This model was intended  for study the linear,  adiabatic
and radial  pulsations  of AGB stars.  The  purpose of this  paper is to
analyze in depth the linear oscillator proposed by these authors, and to
extend  their study to more  massive and  luminous  stars.  The paper is
organized as follows.  In section II, we summarize the basic assumptions
of the model and we mathematically  describe the oscillator,  whereas in
section III we  characterize  its  dynamics  and  describe a sequence of
bifurcations  which occur as the parameters of the model are varied.  In
section  IV  we  compare  our  results  with  those  for  the  perturbed
oscillator, in order to get a better physical  insight.  In section V we
discuss our results and, finally, in section VI we draw our conclusions.

\section{Description of the Model}

In the search for a description which embodies the essentials of stellar
oscillations,  we  follow  the  simple  model  of the  driven  keplerian
oscillator  derived in  \cite{IFH92},  and  successfully  applied to AGB
stars.  The model assumes that the compact stellar interior is decoupled
from the extended  outer layers.  The driving  originates in the stellar
interior  and  consists of a  pulsation  generated  by  pressure  waves.
Moreover, we consider  only the case of  sinusoidal  driving,  where the
outer layers are driven by the interior pressure waves that pass through
a  transition   zone   characterized   by  a  certain   coefficient   of
transmission.  The  motion  is  calculated  at   successive   states  of
hydrostatic  equilibrium.  No back  reaction of the outer  layers on the
inner ones is considered.

We denominate the driving  oscillator  ``the  interior''  and the driven
oscillator  ``the  mantle''.  These are  separated by a transition  zone
through which the pressure waves from the interior  propagate until they
hit the mantle and dissipate.  The driving  oscillator is represented by
variations of the interior  radius,  $R_{\rm c}$, around an  equilibrium
position,  $R_0$,  according  to:  $R_{\rm  c}=  R_{0}+\alpha  R_{0}\sin
\omega_{\rm c} \tau$ where $\alpha$ and $\omega_{\rm c}$ are the fractional
amplitude and the frequency of the driving, respectively.  The mantle is
represented  by a single  spherical  shell of mass $m$ at  instantaneous
radius  $R_{\rm m}$.  In absence of any driving  force, the  equation of
motion is given by:

\begin{equation}
\frac{d^2 R_{\rm m}}{d\tau^2}=\frac{4\pi R_{\rm m}^2}{m}P
                             -\frac{GM}{R_{\rm m}^2},
\end{equation}

\noindent  where $\tau$ is the time, $P$ is the pressure  inside $R_{\rm
m}$, and $M$ is the  mass of the  rigid  core.  As in  \cite{IFH92},  we
assume  that  the gas  follows  a  polytropic  equation  of  state  with
$\gamma=5/3$.  If the region where the pulsation occurs is assumed to be
nearly  isothermal  (i.e.,  $R_{\rm  m}\approx  R_{*}$, with $R_{*}$ the
equilibrium  radius of the star), the equation of motion near the mantle
reduces to

\begin{eqnarray}
\frac{d^2 r}{dt^2}&=&-\frac{1}{r^2}\left(1-\frac{1}{r}\right) \nonumber\\
&&+ Q\omega ^{4/3}\alpha\sin[\omega(t-r+r_0+\alpha r_0\sin\omega t)]
\end{eqnarray}

\noindent where we have  introduced the following set of  nondimensional
variables:

\begin{eqnarray}
        r& \equiv  & R_{\rm  m}/R_* \nonumber \\
        r_0&  \equiv &R_0 /R_*\nonumber \\
        t& \equiv& \omega_{\rm  m}\tau\nonumber\\
        \omega& \equiv& \omega_{\rm c} / \omega_{\rm m}= r_0^{-3/2}. \nonumber
\end{eqnarray}

\noindent In this equations

\twocolumn[
\hsize\textwidth\columnwidth\hsize\csname@twocolumnfalse\endcsname
\begin{figure}
\vspace*{11.5cm}
\includegraphics{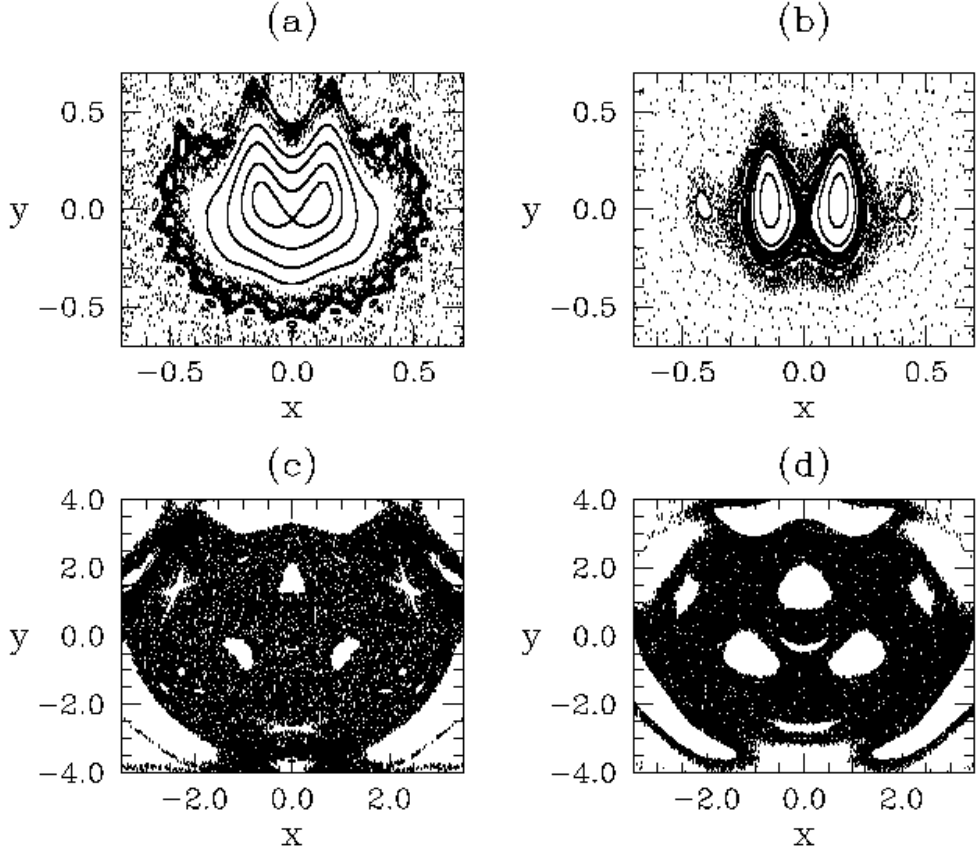}
\caption{Poincar\'{e}  map for Eq.(2.3) with  $\omega=20.1$  (typical of
	AGB stars) and {\sl (a)} $\alpha=0.1$,  {\sl (b)}  $\alpha=0.4$,
	and for  $\omega\simeq 3$ (typical of Super-AGB  stars) and {\sl
	(c)}  $\alpha=0.1$,  {\sl  (d)}   $\alpha=0.4$.  In  all  cases,
	$\epsilon=1$.  Note the difference in the scales of the axes.}
\label{fig1}
\end{figure}
]

$$\omega_{\rm m}\equiv \Big(\frac{GM}{R_*^3}\Big)^{1/2}$$

\noindent is the characteristic  dynamical  frequency of the system and $Q$
the  transmission  coefficient of the transition  zone through which the
pressure waves from the interior  propagate.  In the limit  $r\approx 1$
(small  oscillations) and absorbing some terms into $t$ as phase shifts,
one gets:

\begin{equation}
\frac{d^2 x}{dt ^2} =- x+ \epsilon \sin [ \omega ( t -x+ \alpha \omega^{-2/3}
\sin \omega t )]
\end{equation}

\noindent  where  $\epsilon=Q\omega  ^{4/3}\alpha$  is the total driving
amplitude,  and $x\equiv  r-1$.  The parameter  $\omega$ is a measure of
the core/envelope ratio, and provides information on the location of the
source  of  the  driving  in  the   stellar   interior.  Note  that  for
$\epsilon=0$,  Eq.(2.3)  transforms  into the classical  equation of the
linear  oscillator,  $\ddot x=-x$.  All the interesting  features of the
motion  are  generated  by the  perturbation  (the  second  term  of the
right-hand side of Eq.(2.3) with $\epsilon \neq 0$) and its  interaction
with  the  unperturbed  motion.  The  system  we are  dealing  with is a
periodic--time dependent Hamiltonian system.

\begin{eqnarray}
H(x,y;t)&=&\frac{x^2+y^2}{2} \nonumber\\ &&-
\frac{\epsilon}{\omega} \cos
         [\omega(t-x+\alpha\omega^{-2/3} \sin \omega t)]
\end{eqnarray}

\noindent The associated  Poincar\'{e}  map is an area  preserving  map.
Hence, its dynamics does not exhibit attractors or repellers.

\section{Characterization of the oscillator}

The study performed in \cite{IFH92}  explored a limited set of values of
the  parameter  space:  $\epsilon=0.5,  \, 0.75, \, 1.0$,  and  $\alpha=
0.1,\, 0.2,\, 0.4 $ for values of $\omega$  around $\simeq 20$, which is
characteristic  of regular AGB stars ($M \le 8 \, M_\odot$).  Their main
results and conclusions  are that small values of $\omega$  produce more
strongly chaotic pulsations and that for large values of $\alpha$ stable
orbits in some  definite  regions of the phase space are  obtained.  Our
aim is to extend  the  previous  study to  Super-AGB  stars,  which  are
characterized  by a  different  value  of  $\omega$.  For  the  physical
conditions found in Super-AGB stars,  $8\,M_\odot \le M\le  11\,M_\odot$
and $R_*\sim 450\,  R_\odot$, it turns out that  $\omega\ga 3$.  We will
also explore a wider range of the parameter space than that investigated
in \cite{IFH92}.  Contrary to  \cite{IFH92},  who adopted a fourth-order
predictor-corrector scheme for the numerical integration of Eq.(2.3), we
use a fourth order  Runge-Kutta  integrator  with step-size  control and
dense output as described in \cite{HEA93}.  We have tested several other
integrators,  specifically designed for stiff problems \cite{SG}, and we
have  obtained the same  results for a given set of initial  conditions.
Thus we  conclude  that our  numerical  integrator  is  appropriate  for
problem under study.

\begin{figure}
\vspace*{14.7cm}
\includegraphics{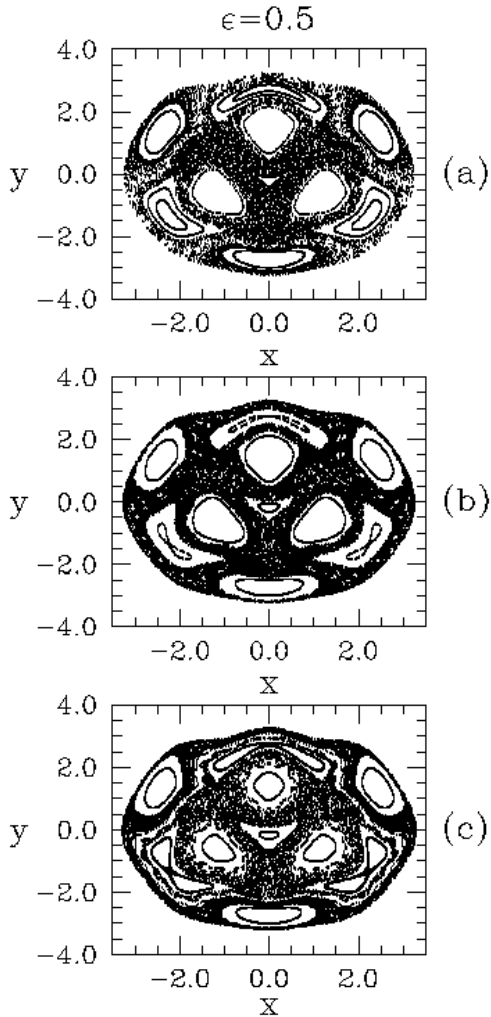}
\caption{The  effect of  increasing  $\alpha$ on the  Poincar\'{e}  map:
     {\sl (a)}  $\alpha=0.2$;  {\sl (b)}  $\alpha=0.3$ and {\sl (c)}
     $\alpha=0.4$.  $\epsilon=0.5$ in all cases.}
\label{fig2}
\end{figure}

\subsection{AGB stars vs. Super-AGB stars: the role of $\omega$}

In order to compare the behaviour for two  different  values of $\omega$
(and, thus, the  differences  between  pulsations  of AGB and  Super-AGB
stars),  Figure 1 shows the  Poincar\'{e}  maps  obtained  with  $\omega
\simeq  20$,  typical of regular  AGB stars  (panels a and b), and those
obtained for  $\omega\simeq  3$ typical of Super-AGB stars (panels c and
d),  for  the  same  initial   conditions.  As  $\omega$  increases  the
behaviour  becomes more  irregular, and the islands and the structure of
panels  (a)  and  (b)  quickly  disappear,  leading  to a  more  chaotic
behaviour.  Consequently,  we expect that the  pulsations  of  Super-AGB
stars will exhibit a more chaotic behaviour than those of AGB stars.

The  phase   portrait   of  our  model  is  a  typical   example  of  an
area-preserving   map.  For  very  small  values  of  the   perturbation
parameter,  it  exhibits  invariant  circles  on  which  the  motion  is
quasi--periodic.  Three periodic island chains  surround the fixed point
of the map.  For large perturbations, because of the breaking--up of the
invariant  circles  and  splitting  of the  separatrices  of  hyperbolic
periodic  points,  chaotic  zones are present, as well as islands in the
stochastic  sea.  Until  recently, it was thought that the  existence of
islands was not very important in  determining  the origin and character
of chaos in  Hamiltonian  systems.  However, it has been recently  shown
that   the   boundaries   of   such   islands   are   very   interesting
\cite{Z99,ZEA97,Z98}.  In particular,  crossing an island boundary means
a transition from a regular  periodic or  quasiperiodic  behaviour to an
irregular   one   that   lies  in  the   stochastic   sea.  One  of  the
particularities  of this singular zone is illustrated in Figure 1a.  One
can see a central island embedded in the domain of chaotic motion with a
boundary separating the area of chaos from the region of regular motion.
By changing a control  parameter  of the  system,  several  bifurcations
occur which  influence the topology of the boundary zone, leading to the
appearance  and  disappearance  of smaller  islands.  This may result in
self-similar  hierarchical  structures  of islands  which is crucial for
understanding  chaotic  transport in the stochastic  sea and the general
dynamics, since they represent high-order resonances.

Figure 1b  illustrates  another  essential  phenomenon  typical  of some
Hamiltonian  systems.  It is linked to the  existence of a complex phase
space topology in the  neighborhood  of some islands:  a trajectory  can
spend an  indefinitely  long time in the  boundary  layer of the  island
(i.e., a {\sl dynamical trap}).  The main uncertainty in  characterizing
this  phenomenon   concerns  the  level  of  stickiness  (that  is,  its
characteristic  trapping  time) which depends on the  parameters  of the
system  in  a  still  unknown  manner.  In  general,  the   hierarchical
structure of the islands in the boundary  zone can explain the origin of
the stickiness of the trajectory within this region, which is the reason
why  this   behaviour   is  called  {\sl   hierarchical--islands   trap}
\cite{Z98}.  Much  attention was  generally  paid to  understanding  the
structure of these  islands  because  different  islands  correspond  to
different physical processes responsible for their origin.  The boundary
layer of an island may include  higher-order  resonant islands.  Another
important  reason  is that  the  topology  of the  islands  could  be an
indicator  for  the  vicinity  of   bifurcations.  However,  no  general
description  for the birth and collapse of  hierarchical  islands exists
yet.

\twocolumn[
\hsize\textwidth\columnwidth\hsize\csname@twocolumnfalse\endcsname
\begin{figure}
\vspace*{11cm}
\includegraphics{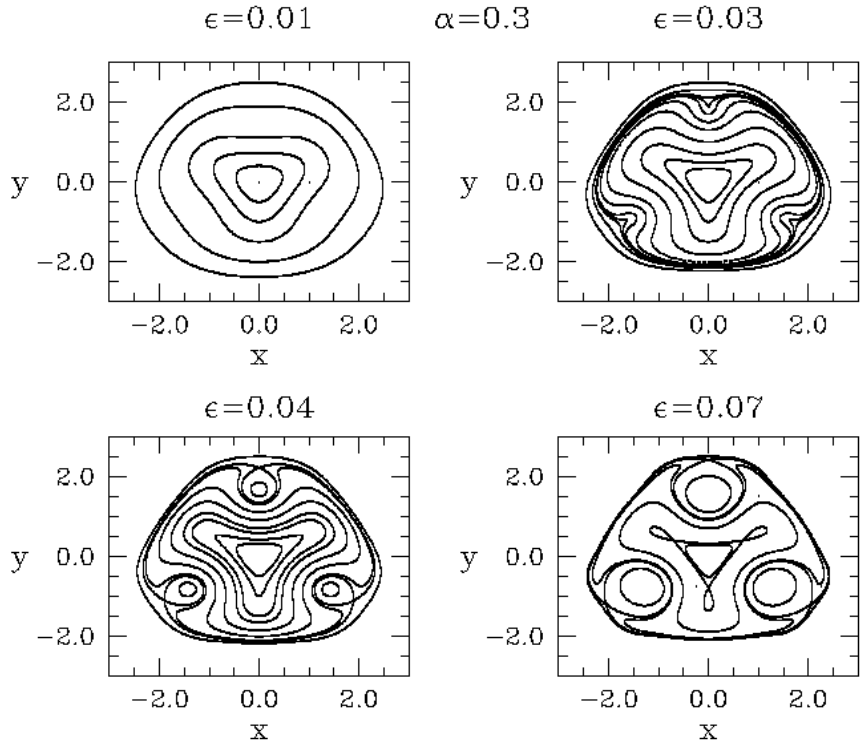}
\caption{Poincar\'{e}  maps for  Eq.(2.3),  $\alpha=0.3$  and  different
	 values of  $\epsilon$.  The  successive  birth of two dimerized
	 island chains.  The  homoclinic  and  heteroclinic  connections
	 appear in red color.}
\label{fig3}
\end{figure}
]

\subsection{The role of the fractional driving amplitude, $\alpha$}

In order to  characterize  in depth the behaviour of our system, we have
performed  a  thorough   parametric   study  by  varying   $\alpha$  and
$\epsilon$,  while  keeping  $\omega$  constant at the value  typical of
Super-AGB  stars.  Reasonable   values  of  the  fractional   amplitude,
$\alpha$,  range from 0.1 to 0.4, whereas the total  driving  amplitude,
$\epsilon$,  varies  between 0.1 and 1.0.  Since  $\alpha$  is the ratio
between the  amplitude  of the  internal  driving  and the radius of the
star,  the  upper  limit  considered  in this  study is  40\%,  which is
physically   sound.  In  general,  a  star  is   characterized   by  its
compactness,   $\omega$,  and  the  interior-mantle   coupling  strength
$Q=\epsilon  /(\omega^{4/3}\alpha)$, which should stay within the limits
of 0 (0$\%$  transmission)  and 1  (100$\%$  transmission)  in  order to
maintain its physical meaning.  Therefore some specific  combinations of
$(\epsilon,  \alpha)$  that  lay  outside  this  range  for $Q$  must be
regarded with caution,  since they have no physical  meaning.  According
to all these  considerations,  we have  investigated the dynamics of the
system associated to Eq.(2.3) rewritten below in the form of a perturbed
oscillator,

\begin{eqnarray}
\dot{x}&=&y\nonumber\\
&&\\
\dot{y}&=&-x+\epsilon \sin [\omega (t-x+b\sin\omega t)]\nonumber
\end{eqnarray}

\noindent where $b=\alpha\omega^{-2/3}$, maintaining a constant value of
$\omega\simeq 3$, as discussed above.

In a first step, we have studied the qualitative changes in the dynamics
(i.e.,  bifurcations)  as the  parameters  $\epsilon$  and  $\alpha$ are
varied.  Poincar\'{e}   maps   corresponding   to  the  case  $\alpha\in
(0.1-0.4)$ and $\epsilon=0.5$, for several initial conditions, are shown
in  Figure  2.  The  maps  are   characterized  by  the  same  geometric
structure:  a region, centered around  $(x,y)=(0,0)$,  of closed orbits,
surrounded  by a region of chaotic  orbits.  As it can be seen in Figure
2, as  $\alpha$  increases,  the  central  region of  regular  behaviour
expands to the detriment of the stochastic sea.

We have also  analyzed the  behaviour  of the system for {\sl  negative}
values of $\alpha$.  A quick look at the driving  allows one to see that
negative values of $\alpha$  correspond to a mere change of phase in the
driving and, therefore, correspond also to physically  meaningful cases.
However, and for the sake of clarity and  conciseness,  in this paper we
will  concentrate our efforts on the positive  domain of this parameter.
Nevertheless,  we would  like to  remark  at this  point  that the phase
portrait  of the  Poincar\'{e}  map  changes  drastically  depending  on
whether $\alpha$ is positive, negative or almost zero.

\twocolumn[
\hsize\textwidth\columnwidth\hsize\csname@twocolumnfalse\endcsname
\begin{figure}
\vspace*{15cm} 
\includegraphics{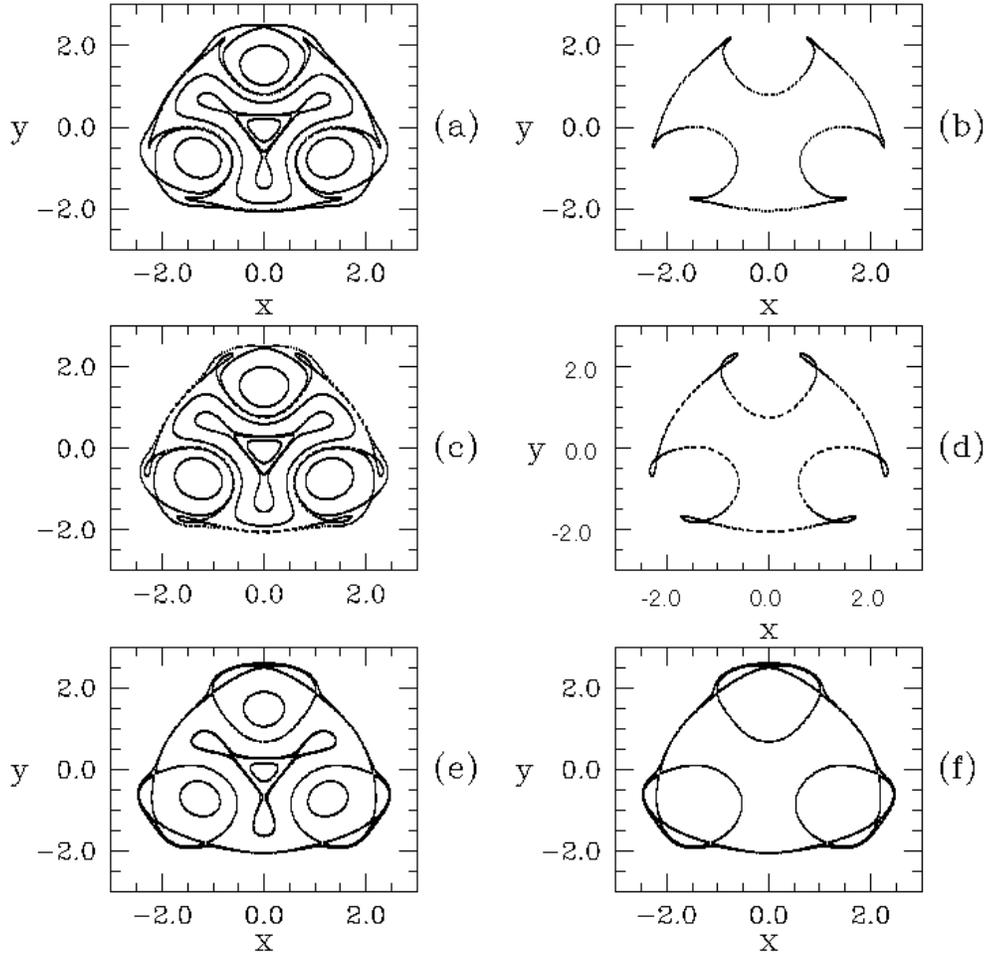} 
\caption{Creation  by  saddle--  center  bifurcation  of a pair of three
	 periodic  orbits  for  $\alpha=0.3$  and  increasing  values of
	 $\epsilon$.  {\sl   (a):}  $\epsilon=0.0875$   ---  the   phase
	 portrait at the threshold of the  bifurcation;  {\sl (b):}  the
	 invariant circle with 6 cusps for the case {\sl (a)}; {\sl (c)}
	 $\epsilon=0.095$  --- two  intertwined  orbits of period  three
	 have been created,  inside each new-born loop there is a period
	 three  elliptic  point; {\sl (d):}  a sequence of  heteroclinic
	 and  homoclinic  connections  between  the new born  hyperbolic
	 points;  {\sl (e):}  $\epsilon=0.11725$  --- the  threshold  of
	 global   bifurcation;  {\sl  (f):}  the  new  topology  of  the
	 separatrices.}
\label{fig4}
\end{figure}
]

\subsection{The role of the total driving amplitude, $\epsilon$}

Next we have focused on values of  $\epsilon  \ll 1$ in order to observe
in detail the departure of our equation from the harmonic  oscillator as
this parameter  increases.  We restricted the study of the  Poincar\'{e}
map to a rectangle  limited by initial  conditions  close to  reasonable
values of the radius and  velocity of the  mantle.  The  restriction  is
also  due to the  approximations  involved  in  deriving  Eq.(2.3).  For
$\epsilon = 0$ we get an integrable  Hamiltonian system, whose integrals
of  motion  are  the  tori  given  by  the  condition  $  x^2  +  y^2=C,
~t=[0,2\pi/\omega)$.  The Poincar\'{e}  map has the elliptic fixed point
(0,0), which  corresponds to a periodic orbit of period  $P=2\pi/\omega$
of the perturbed Hamiltonian system.

We have chosen the range $\epsilon\in  (0,0.12)$ and $\alpha=0.3$ and we
have obtained a cascade of local and global bifurcations.  Note that the
considered pairs of parameters $(\epsilon,\alpha)$ are located above the
curve  $\alpha^*(\epsilon)$,  which is the curve of triplication  of the
elliptic fixed point.  According to \cite{DEA00,AEA88} the corresponding
Poincar\'{e}  map is in these  cases a nontwist  map.  The  sequence  of
detected  local  and  global  bifurcations  is  typical  for  such a map
\cite{CNEA96,CNEA97,S98,P01}.  Figure 3 illustrates  the birth in stages
of two dimerized  island  chains  containing  periodic  points of period
three.  This is a typical scenario of creation of new orbits in nontwist
maps.  The  upper-left   panel  exhibits  the  phase   portrait  of  the
near--to--integrable  map.  The elliptic  fixed point is  surrounded  by
invariant  circles.  At  $\epsilon=0.03$  starts  the birth of the first
dimerized island chain, namely a regular invariant circle  bifurcated to
an invariant  circle with three cusps.  The newly born dimerized  island
chain is clearly seen in the bottom--left panel of the Figure 3, where a
new circle with three cusps,  closer to the the fixed point can be seen.
Both dimerized island chains are illustrated in the bottom--right panel.
Between them the invariant circles are meanders.  Note that each cusp is
a point of saddle--center creation \cite{HK91}.  This scenario is a good
example to illustrate the theoretical results concerning the creation of
a twistless circle after triplication \cite{DEA00,AEA88}.

\begin{figure}
\vspace*{14.5cm} 
\includegraphics{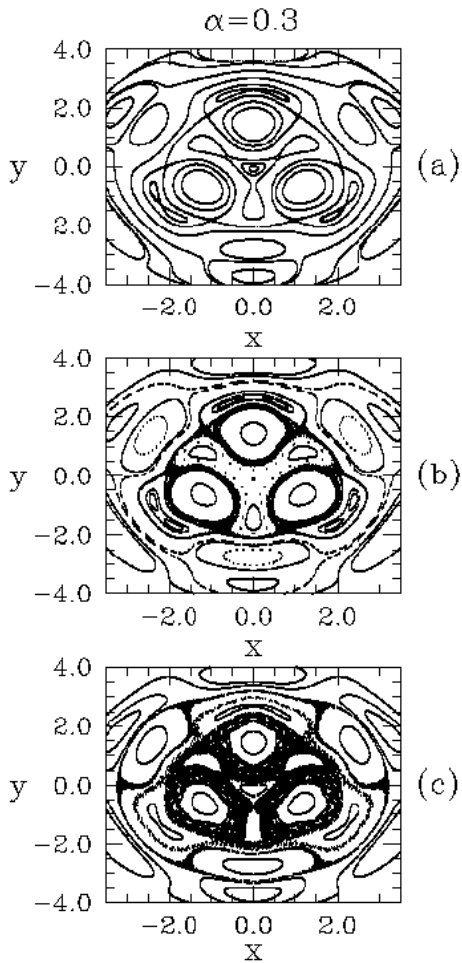} 
\caption{Poincar\'{e}  sections for $\alpha=0.3$ and different values of
	$\epsilon$:  {\sl (a)} $\epsilon=0.2$; {\sl (b)} $\epsilon=0.3$,
	and  {\sl  (c})  $\epsilon=0.4$.  In red and  green  colors  are
	represented  evolutions of initial conditions close to the first
	and respectively second separatrix.}
\label{fig5}
\end{figure}

Around the last born three--periodic  dimerized island chain, a sequence
of local and global bifurcations  occurs.  This is illustrated in Figure
4, where it can be seen that two  independent  orbits of the same period
three are created by cusp  bifurcation.  They evolve in such a way that,
finally,  interact  with the orbits which belong to the first  dimerized
island chain.  As $\epsilon$  increases the newly born  elliptic  points
approach the hyperbolic  points of the dimerized chain.  When $\epsilon$
reaches  a value of  0.11725  a global  bifurcation  occurs:  the  newly
created orbits  interfere,  and the  hyperbolic  points of the dimerized
island chain become hyperbolic points with homoclinic  eight-like orbits
encircling the new created  elliptic  points.  We have also noticed that
as  $\epsilon$  increases,  the process of  creation  of three  periodic
orbits repeats outside the region of the  Poincar\'{e}  map we focus on,
and the new orbits form chains of vortices, not dimerized island chains.
This  behaviour  is due to the  oscillating  character  of the  nontwist
property of the perturbation.

In a second step we have paid attention to larger values of  $\epsilon$.
In particular  we have studied the range  $\epsilon\in  (0.2,0.5)$.  The
increase  in  strength  of  the  external   perturbation   destroys  the
separatrices  (Figure  5,  top)  by  clothing  every  one of  them  in a
stochastic  layer  (Figure 5, middle).  As the  thickness  of the layers
increases  with the  perturbation,  depending  on the  positions  of the
separatrices  in the phase space, they can merge forming the  stochastic
sea (Figure 5, bottom).

There are many islands, which the chaotic  trajectory  cannot penetrate.
Within an island there are  quasiperiodic  motions  (invariant tori) and
regions of trapped  chaos.  The  stronger  the chaos is, the smaller the
islands  and the larger the  fraction  of phase  space  occupied  by the
stochastic sea.  The coexistence of regions of regular dynamics  (closed
orbits) and regions of chaos in the phase space is a  wonderful  example
of the property  which  differentiates  chaotic  systems  from  ordinary
random processes, where no stability islands are present.  This property
makes possible the analysis of the onset of chaos and the  appearance of
minimal  regions  of chaos.  To  summarize,  the  system  undergoes  the
following bifurcations, as it results from Figures 3, 4 and 5:

\begin{enumerate}
\itemsep=-5pt

\item [(a)] For   $\epsilon\in[0,0.03)$,   the  phase  portrait  of  the
	    Poincar\'{e}   map  is  similar  to  that  of  the  harmonic
	    oscillator,  with the  elliptic  fixed  point  $(x,y)=(0,0)$
	    surrounded by almost circular closed orbits.

\item [(b)] For $\epsilon \in  [0.03,0.04)$,  in addition to the central
	    elliptic  point, a dimerized  island chain is born,  namely,
	    three periodic elliptic and hyperbolic points, each elliptic
	    point  being  surrounded  by  a  homoclinic   orbit  to  the
	    corresponding  hyperbolic  point,  and  distinct  hyperbolic
	    points being connected by heteroclinic orbits.

\item [(c)] For $\epsilon \in  [0.04,0.07)$, a second  dimerized  island
	    chain is created, which is a typical configuration after the
	    triplication of an elliptic fixed point.

\item  [(d)] For  $\epsilon  \in  [0.07,0.2)$,  a sequence  of local and
	     global bifurcations  occurs which interphere with the first
	     created   dimerized   island   chain:  a  regular   meander
	     bifurcates  to a  meander  with 6  cusps.  A new  dimerized
	     island chain is created by saddle--center  bifurcation.  If
	     we denote the cusps as $1,2, \ldots, 6$ in clockwise order,
	     then the elliptic and  hyperbolic  points born from the odd
	     cusps, form independent  orbits of those born from the even
	     cusps.  This particular  dimerized  island chain approaches
	     the original one, and as a result the hyperbolic  points of
	     the two chains are connected  giving rise to a pattern of a
	     complicated topological type (Figure 4f).

\item [(e)] For  $\epsilon  \in  [0.2,0.3)$,  we notice a more  definite
	    chaotic   behaviour   created  by  the  instability  of  the
	    remainings of the first  heteroclinic  orbit  engulfing  the
	    periodic points of case (c).

\item [(f)] For $\epsilon \in [0.3,0.5]$, we notice the extension of the
	    chaotic  orbits in the  central  region,  together  with the
	    continuous  generation  of pairs  of  three  elliptic  fixed
	    points in the outer region as $\epsilon$ increases.

\end{enumerate}

\begin{figure}
\vspace*{10.5cm}
\includegraphics{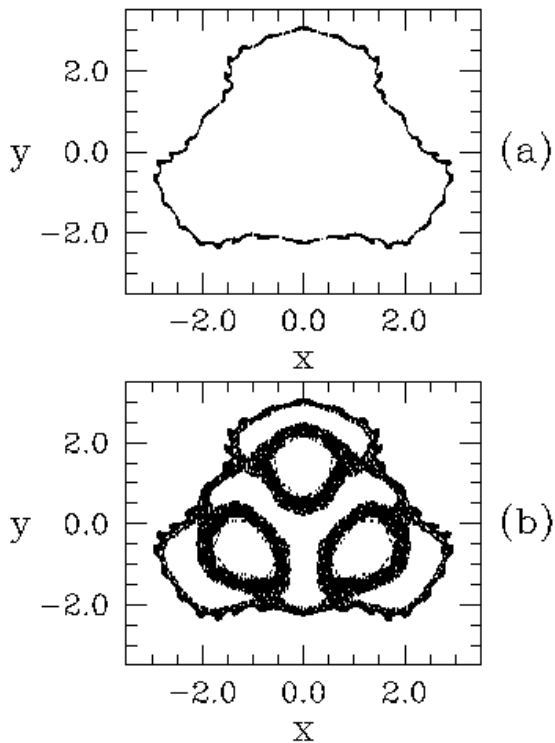}
\caption{{\sl  (a):}  A meander near an  invariant  circle with 35 cusps
         for  $\epsilon=0.4$  and $\alpha=0.3$;  {\sl (b):}  two orbits,
         one filling the meander and the  second, a chaotic  orbit,
         confined by the meander.}
\label{fig6}
\end{figure}

Another  important  feature  of the  phase  portrait  characteristic  of
nontwist  maps is the  existence  of  {\sl  meanders}  (i.e.,  invariant
circles which fold exactly as a meander).  Meanders are created  between
two successively born dimerized island chains, and between two chains of
vortices.  In twist  standard--like maps they become usual circles after
the  reconnection  of  the  two  chains.  Note  that  in  our  case  the
reconnection of the first two created  dimerized island chains, does not
occur,  and as  $\epsilon$  increases  the  meanders  break--up  leaving
instead a Cantor set.  The top panel of Figure 6  illustrates  a meander
near a pair of dimerized  island chains  containing  periodic  orbits of
period 35.  Meanders  appear to be robust  invariant  circles, even when
the nearby orbits are chaotic  (Figure 6, bottom panel).  This behaviour
was observed in nontwist  standard-like  maps  \cite{S98}, but until now
there is no explanation for this robustness.

\section{Comparison with the perturbed oscillator}

In order  to get a  better  insight  of the  underlying  physics  of the
oscillator  studied so far, in this  section we are going to  compare it
with the  motion  of a  perturbed  oscillator,  which  has been  already
studied  extensively.  This is  important  since,  as it will  be  shown
below, the formal appeareance of Eq.(2.3) does not differ very much from
that of a  perturbed  oscillator.  In order to make this clear  consider
the motion of a perturbed linear oscillator in the form of:

\begin{equation}
\ddot{x} +\omega_0^2 x=\epsilon \sin[\omega x-\omega t]
\end{equation}

The left-hand side of Eq.(4.1)  describes very simple dynamics of linear
oscillations of frequency  $\omega_0$.  To ease the comparison  with the
perturbed oscillator, Eq.(2.3) can be rewritten as:

\begin{equation}
\ddot{x}+\omega_0^2 x= -\epsilon [\omega x - \omega (t)t],
\end{equation}

\noindent where

\begin{equation}
\omega(t)=\omega \left( \frac{b\sin\omega t}{t} + 1 \right)
\end{equation}

Both equations share common features.  It is important to point out here
that for our system  $\omega_0=1$,  as it  results  from  Eq.(2.3).  The
Hamiltonian associated with the system given by Eq.(4.1) is:

\begin{equation}
H= \frac{(\dot{x}^2+\omega_0^2x^2)}{2}- \frac{\epsilon}{\omega}
\cos[\omega x-\omega (t) t]
\end{equation}

Consequently, for weak  perturbations the results obtained so far should
be similar to the case of a  perturbed  oscillator.  And indeed  this is
the case, the only difference  resides in the period 3 homoclinic orbit,
that is in point d) of the summary given in section III.C.  The creation
of this period-3  homoclinic  structure was  illustrated  in Figure 2 as
$\alpha$  increases from  $\alpha=0.2$  to  $\alpha=0.3$.  This suggests
that the ultimate reason of this difference might be the presence of the
function $\omega (t)$ from Eq.(4.3) as a weak detuning of resonance.

\twocolumn[
\hsize\textwidth\columnwidth\hsize\csname@twocolumnfalse\endcsname
\begin{figure}
\vspace*{7.0cm}
\includegraphics{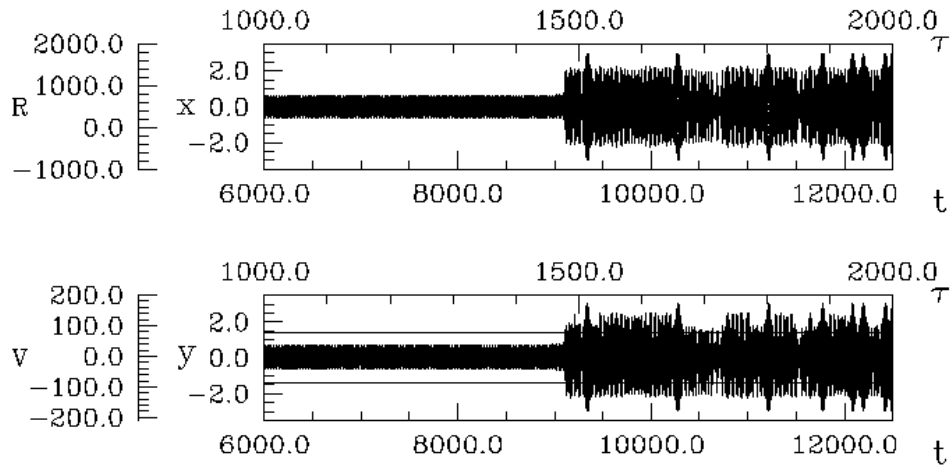}
\caption{Variations of radius {\sl (top)} and velocity {\sl (bottom)} as
         a  function  of  time  for  $\omega\simeq   3$,   $\alpha=0.3$,
         $\epsilon=0.5$   and  the  initial   condition   $(x_0,y_0)   =
         (0.0,0.02)$.  The escape  velocity  of the model star  ($v_{\rm
         esc} \simeq 86$~km/s) is represented as a red line.}
\label{fig7}
\end{figure}
]

We rewrite Eq.(4.2) with $\alpha\neq 0$ in the following way:

\begin{equation}
\ddot{x}+\omega_0^2 x=\epsilon \sin\omega(t-x+b\sin\omega t)]
\end{equation}

\noindent For $\alpha=0$, Eq.(4.5) transforms into:

\begin{equation}
\ddot{x}  +\omega_0^2 x=\epsilon\sin[\omega(t-x)]
\end{equation}

\noindent All these ordinary  differential  equations, together with the
well  known  equation  for  the  perturbed  oscillator,  have  the  same
mathematical structure.  It is an example of {\sl weak chaos}, where the
perturbation   itself  creates  the  separatrix  network  at  a  certain
$\epsilon_0$  and  then  destroys  it  as  $\epsilon$  increases  beyond
$\epsilon_0$  by  producing   channels  of  chaotic   dynamics.  For  an
unperturbed  equation, the unperturbed  Hamiltonian $H_0$  intrinsically
has  separatrix  structures  and the  perturbation  clothes them in thin
stochastic layers, an example of {\sl strong chaos}.  In the phase space
there appear invariant curves (deformed tori) embracing the center which
do not allow  diffusion in the radial  direction.  Inside these cells of
the web, motion occurs along  closed-orbits,  around the elliptic points
from the  centers of the cells.  With the  increase  of  $\epsilon$  and
creation  of the  stochastic  layers,  particles  can  wander  along the
channels  of the  newly  born  web,  a  phenomenon  that  represents  an
universal  instability  and gives  birth to  chaotic  fluctuations.  The
heteroclinic  structures  formed as the perturbation  increases  through
this bifurcation are what  differentiates  our equation from the typical
equations  of  Hamiltonian  chaos and in  particular  from  those of the
perturbed oscillator.

\section{Astrophysical interpretation of the results}

In order to provide the reader with a feeling of the physical ranges for
the stellar  fluctuations  in radius and velocity, in Figure 7 we show a
particular time series of the model presented above.  In this figure the
variations  of radius and velocity as functions of time are  represented
both in physical units (solar radii, km/s and years,  respectively)  and
in adimensionalized units --- as we have been doing so far.

Generally  speaking, the primary  information that one can derive from a
generic computed or observed time series is the spectral distribution of
energies  (or  amplitudes)  of the light  curve.  We found that for some
initial  conditions,  the  resultant  time  series  do not show a single
frequency  or a  small  set of  frequencies  but,  in  addition,  linear
combinations of the primary  frequencies may also appear in the spectra.
This  behaviour  has been found in real stars quite  frequently,  and in
particular in LPVs, Cepheids and RR~Lyrae stars \cite{KB00}.

In any  calculation of a  time-varying  phenomenon,  like the one we are
describing  here, there is always the  important  practical  question of
when  the   simulations   should  be   terminated.  For  full  nonlinear
hydrodynamic simulations, even in the case in which a stable limit cycle
seems to have set in, one often is worried  about the fact that  thermal
changes,  which  take  place on a  longer  timescales,  could  still  be
occurring.  In a recent paper  \cite{YT96}, the authors have carried the
calculations of Mira variable  pulsations  much further than in previous
investigations  and, to their  surprise, they have found that the actual
behaviour  was  significantly  different  from  what  it was  previously
thought.  To be precise,  they  obtained a new modified  ``true''  limit
cycle which is quite different from the earlier,  ``false'' limit cycle.
In agreement with these results, the numerical integration of our system
presents a similar  behavior,  despite the very crude  approach  adopted
here which results in an extreme simplicity of the model.  This is shown
in Figure 7 for a particular orbit:  during the first $\sim 1400$ yr the
star oscillates quite regularly, and with small  amplitudes.  After this
period of time, the quiet phase  suddenly  stops and  pulsations  become
more violent and chaotic.  Moreover,  at late times the  velocity of the
outer layers  exceeds the escape  velocity and, hence, mass loss is very
likely to occur, in accordance with the  observations of LPVs, which are
the observational counterparts of Super-AGB stars.

\begin{figure}
\vspace*{14cm}
\includegraphics{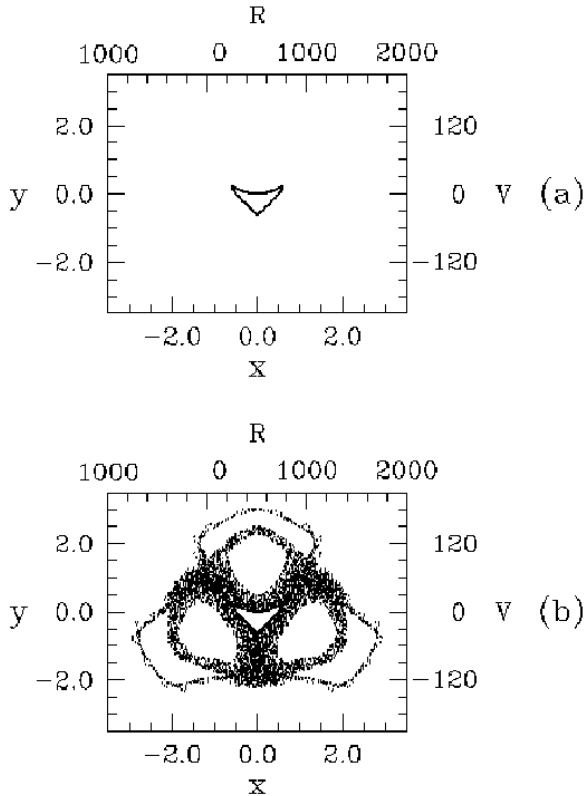}
\caption{Poincar\'{e}  map for the orbit of Figure  7.  Panel  {\sl (a)}
        includes  the  points  $R(\tau)$  and  $V(\tau)$  for the  time
        interval  (1000-1400)  years, whereas panel {\sl (b)} shows the
        points   $R(\tau)$   and   $V(\tau)$   for  the  time  interval
        (1000-2000) years.}
\label{fig8}
\end{figure}

This behaviour --- a quiet phase, followed by an extremely violent phase
--- is a typical case of a dynamical trap discussed in section III.  The
Poincar\'{e}  map associated with the radius and velocity  variations in
Figure 7 is shown  in  Figure  8.  None of  these  figures  include  the
oscillations  previous to 1000 years because the behaviour is similar to
the one of the time interval  between 1000 and 1400 years.  This example
corroborates  the suggestion,  first proposed by \cite{BYP77}  and later
found in the very detailed  numerical  calculations of \cite{YT96}  that
the long-term  effects in Mira variables  have an important  role in our
understanding of the mechanism which drives mass-loss.  This result also
argues in favour of the  capability  of simple  models  to  capture  the
underlying  dynamics of the  physical  system.  To put it in a different
way, our model, despite its simplicity,  reproduces  reasonably well the
results  obtained with full,  sophisticated  and nonlinear  hydrodynamic
models.  It is  nevertheless  important  to realize  here that our model
does not incorporate the secular effects induced by the thermal  changes
and,  hence,  it  is  quite  likely  that  this  kind  of  behaviour  is
intrinsically  associated  with  the  physical  characteristics  of  the
oscillations of real stars.

In  order  to get a  more  precise  physical  insight  a  time-frequency
analysis would be very  valuable.  Consequently  we devote the remaining
of this  section to such a purpose.  Most often,  astronomers  have used
wavelet  transforms  in  their   investigations  of  the  time-frequency
characteristics of variable light curves.  However, \cite{KB97} compared
the results  obtained with different  time-frequency  methods using both
real light-curves and synthetic  signals.  From these tests they firstly
concluded  that the G\'{a}bor  transform  \cite{G46}  provides much more
informative  results  on the high  frequency  part of the data  than the
wavelet  transform,  but, on the  contrary,  they  also  found  that the
time-frequency analysis using the Choi-Williams distribution \cite{CW89}
is definitely superior to both methods (at least on the data they used).
Further  investigations  carried out by the same authors \cite{BK00} led
them to conclude  that in general one cannot  claim {\sl a priori}  that
any of these  methods is better than the others.  In fact, this  depends
largely on the nature of the signal and on which  kind of  features  one
tries  to  enhance.  Therefore,  it  is  always   advantageous   to  use
simultanously   several   of   them.  However,   as  the   Choi-Williams
distribution  is generally  accepted to be the most useful among all the
methods, we will center our analysis  using this tool, as it was done in
\cite{BK00}.

The time-frequency  analysis of a model computed with  $\omega\simeq 3$,
$\alpha=0.3$,  $\epsilon=0.5$  and the initial  condition  $(x_0,y_0)  =
(0.0,0.02)$,  is shown in the central panel of Figure 9.  In order to be
used as a visual help this time series has been reproduced in the bottom
panel of this figure.  We have split the time series into two pieces, in
order to better illustrate the stickiness of the oscillator, and we have
computed  the  Fourier  transforms  of  both  pieces  separately.  These
Fourier transforms are shown in the top panels of Figure 9, the left one
corresponds to $\tau<  1400$~yr,  whereas the right one  corresponds  to
$\tau \ge  1400$~yr.  Note the  difference  in the  scales of the power,
which is much  larger  for the right  panel.  As it can be seen in these
panels the power is suddenly shifted from relatively  large  frequencies
to much smaller frequencies.

\twocolumn[
\hsize\textwidth\columnwidth\hsize\csname@twocolumnfalse\endcsname
\begin{figure}
\vspace{14 cm}
\includegraphics{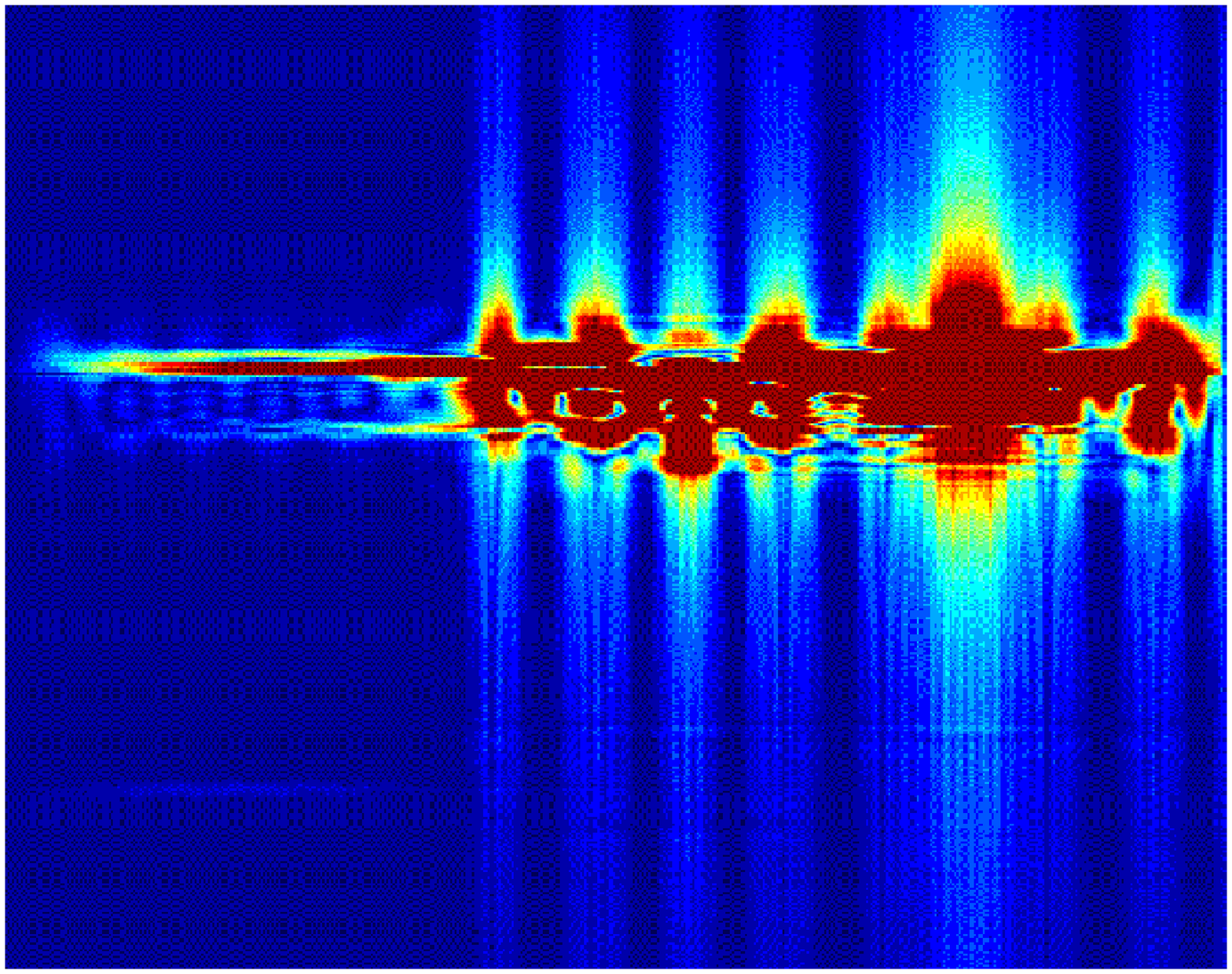}
\includegraphics{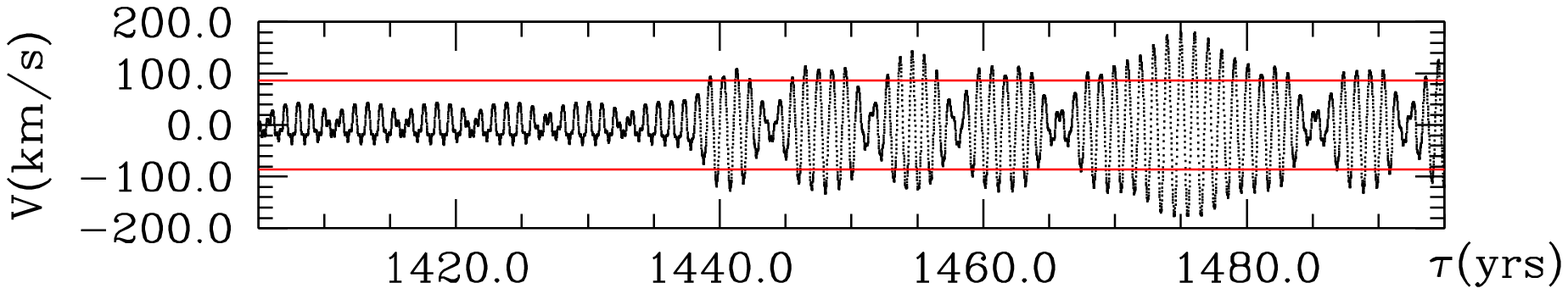}
\includegraphics{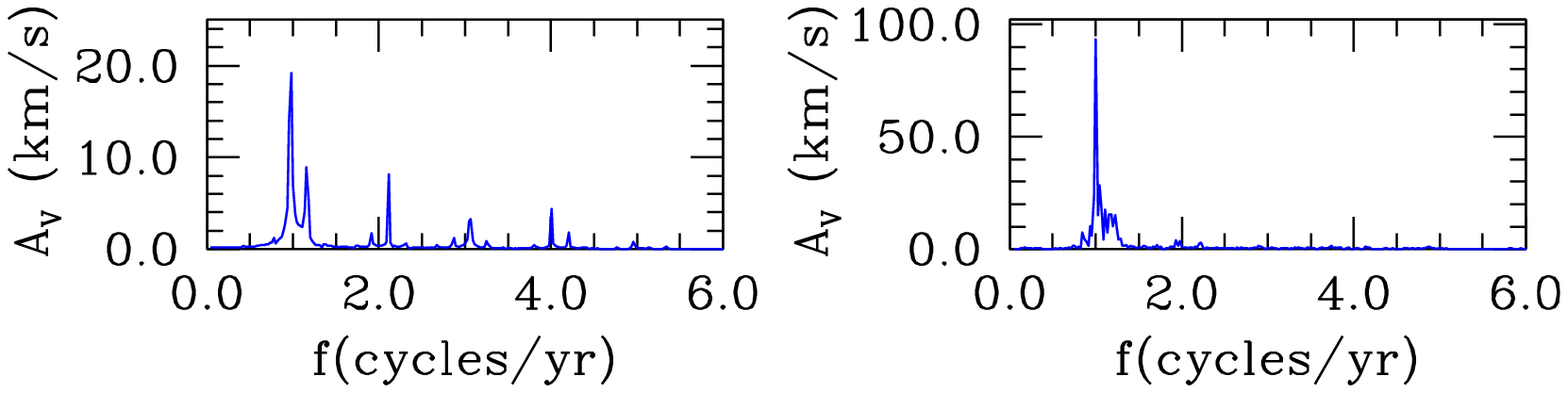}
\includegraphics{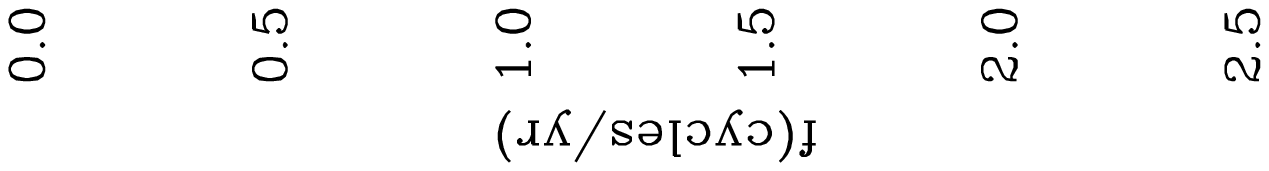}
\caption{Choi-Williams  distribution  of the  velocity  variations  {\sl
	 (central panel)}, Fourier analysis for $\tau<1440$ yr {\sl (top
	 left  panel)}  and for  $\tau  \ge  1440$  yr {\sl  (top  right
	 panel)},  and  the  corresponding   time  series  {\sl  (bottom
	 panel)}.}  
\end{figure} ]

This  behavior  is  even  more  clearly   shown  in  the   Choi-Williams
distribution.  Note as well the  contribution of the frequency  $f\simeq
2.1$  yr$^{-1}$ at small times, just before the  beginning  of the burst
(at $\tau \le  1430$~yr).  Indeed, this small  contribution  could be at
the origin of the  transfer of power to $f\simeq  1.2$~yr$^{-1}$,  which
occurs  inmediately  after $\tau \simeq  1430$~yr  and which  ultimately
leads to the burst at $\tau  \simeq  1440$~yr.  Also  remarkable  is the
small time elapsed  since the  beginning of the transfer of power, which
is only about 10 yr.

\section{Conclusions}

We  have  presented  and  analyzed  the  complex  dynamics  of a  forced
oscillator which is interesting not only from the mathematical  point of
view, but also because it describes with a reasonable degree of accuracy
the main  characteristics of some stellar  oscillations.  This model has
been  previously  used to study the irregular  pulsations  of Asymptotic
Giant  Branch  stars  \cite{IFH92},  and we have  used it to  study  the
pulsations   of  more  massive  and   luminous   stars,  the   so-called
Super-Asymptotic Giant Branch stars.  In doing this we have extended the
previous studies to a range of the parameters  specific for this stellar
evolutionary  phase.  We have found, in agreement with the  observations
of Long Period  Variables which are the  observational  counterparts  of
Super-Asymptotic Giant Branch stars, that the oscillator shows a chaotic
behavior.  It is  important  to realize  as well that this kind of stars
shows a more pronounced  chaotic behavior than regular  Asymptotic Giant
Branch stars of smaller mass and luminosity.  We have also characterized
in depth the full sequence of bifurcations as the physical parameters of
the model  are  varied.  We have  found a rich set of local  and  global
bifurcations  which were not  described  in  \cite{IFH92}.  Among these,
perhaps the most important one is a tripling bifurcation, but meandering
curves,  hierarchical  islands  traps and  sticky  orbits  also  appear.
Correspondingly,  the resulting time series also show a rich  behaviour.
In particular, we have found that although  there are light curves which
show a rather regular  behaviour for certain values of the parameters of
the physical system and given initial conditions, there are as well some
other light curves which show clear beatings or linear  combinations  of
two main  frequencies  up to  terms  of  $2f_0+7f_1$,  being  $f_0$  the
fundamental  frequency  and $f_1$ that of the first  overtone,  and even
more complex orbits.  For the parameters and initial conditions  leading
to more  irregular  behaviour,  we noticed the  existence  of both clear
chaotic  pulsations  and sudden  changes from a  limit-cycle  to chaotic
pulsations, the latter being  associated with the stickiness  phenomenon
characteristic  of  some  Hamiltonian  systems.  For  these  orbits  the
velocity of the very outer layers clearly  exceeds the escape  velocity.
Hence, for these chaotic  pulsations  mass-loss is very likely to occur,
in good agreement with the  observations,  which correlate the degree of
irregularity  with the mass-loss rate.  Regarding the stickiness of some
orbits  which we have found for a set of  parameters,  perhaps  the most
important  result is that the long-term  effects found in real stars are
reproduced  by our model,  despite of its  simplicity,  even  though the
driven  oscillator  studied  here does not  incorporate  the  effects of
secular  changes.  Hence it is quite  likely that this kind of behaviour
which  has  been  already  found  in  full  hydrodynamical   simulations
\cite{YT96},   is    intrinsically    associated   with   the   physical
characteristics  of the  oscillations  of  real  stars  and  not  to the
long-term thermal changes.

\begin{acknowledgements}
This work has been supported by the DGES grant  PB98--1183--C03--02,  by
the MCYT grant  AYA2000--1785, and by the CIRIT grant  1995SGR-0602.  We
also  would  like to  acknowledge  many  helpful  discussions  with G.M.
Zaslavsky.
\end{acknowledgements}

\end{document}